\documentclass[paper,notoc]{JHEP}
\usepackage{epsfig}
\usepackage{amsmath}
\usepackage{amssymb}
\usepackage{cite}

\def\ups{\Upsilon}
\def\qf2{Q^2_{eff}}

\newcommand\jpg[3]{ {{\it J. Phys. } {\bf G #1} (#2) #3} }

\title{Diffractive Photoproduction of $\Upsilon$ at HERA.}

\author{L. L. Frankfurt   \\
  Nuclear Physics Department, \\
  School of Physics and Astronomy, \\
  Tel Aviv University, 69978 Tel Aviv, Israel \\          
  E-mail: \email{frankfur@lev.tau.ac.il}}

\author{M. F. McDermott \\
  Dept. Physics and Astronomy,\\
  University of Manchester, \\
  Manchester, M13 9PL, England \\
  E-mail: \email{mm@a13.ph.man.ac.uk}}

\author{M. Strikman\thanks{current address: Theory Group, DESY, 
Notkestr. 85, 22607, Hamburg, Germany}   \\
  Dept. of Physics, Penn. State University,\\
  University Park, USA \\
  E-mail: \email{strikman@theow01.desy.de}}

  \preprint{DESY 98-196\\
	    MC/TH 98-18\\
            Dec '98\\
	    revised Apr '99}


\abstract{Predictions for the diffractive photoproduction of the 
$\Upsilon$-family at HERA energies, within the 
framework of the analysis by Frankfurt, Koepf
and Strikman, are presented. 
Two novel effects lead to a significant enhancement 
of the original calculation: the non-diagonal (or skewed) kinematics, 
calculated to leading-log($Q^2$) accuracy, 
and the large magnitude of the real part of the amplitude. The 
resultant cross sections are found to agree 
fairly well
with recent preliminary data from ZEUS and H1.
A strong correlation between the  mass of the  diffractively produced state and
the energy dependence of  the cross section is found. In particular, a
considerably stronger rise in energy is predicted than that found in 
$J/\psi$-production.}

\begin{document}

\section{Introduction}
\label{sec:intro}
Diffractive heavy vector meson production was initially  evaluated
to leading-log accuracy in energy \cite{RYSK}, while in \cite{BROD} 
all vector mesons were treated to the leading-log accuracy in $\ln Q^2$
and $\ln 1/x$. It is now well understood in QCD, to leading twist  accuracy
thanks to a generalization of the QCD factorization theorem
to hard exclusive processes \cite{CFS}.
So far, for heavy quarkonium production, 
this knowledge has only been used to confront 
photo- and electroproduction of $J/\psi$ \cite{FKS1,FKS2,RRML}. 
Unfortunately, the light mass of the charm quark leads to 
many theoretical uncertainties in the final result 
(relativistic effects in the wavefunction, 
scale uncertainties, unitarity corrections...).
By studying $\ups$-production, which has recently been observed in 
photoproduction for the first time at HERA \cite{DATA}, 
within the same framework, one may hope to 
pin down some of these uncertainties.

The current paper is an extension of the work presented by Frankfurt et al
\cite{FKS2}. One of the main findings was that the use of 
realistic light-cone wavefunctions for the heavy vector mesons, 
$\psi_V (z,k_t)$, with significant average $k_t$, leads to 
an overall suppression of the cross-section relative to the static, 
$\psi_V (z,k_t) = \delta (z-1/2) \, \delta^{(2)} (k_t) $,  
and Gaussian, $\psi_V (z,k_t) = A \, 
\delta (z-1/2) \, \exp(-a k_t^2 / m_c^2)  $,  
forms  used in  \cite{RRML}.
The hybrid wavefunctions used were designed to interpolate hard QCD behaviour 
at small transverse distances (normalised to the decay width into leptons) 
with quarkonium models at large transverse distances.
In the limit $m_q^2 \rightarrow \infty$ (but $m_q^2 \ll W^2$) 
such Fermi-motion effects of the quarks can be substantiated 
in QCD, because non-quark degrees of freedom will be suppressed 
by the powers of $m_q^2$. Whether such an approximation 
is applicable to the production of states in the $\psi$-family is an open question. 
The $k_t$-suppression was tempered by an enhancement 
due to a rescaling (beyond leading-log($Q^2$)) of the gluon density 
to higher scales, to reflect more accurately the typical transverse 
size of the scattering dipoles.


Here, we present predictions for diffractive (also called exclusive) 
photoproduction of the $\ups$-family using similar hybrid wavefunctions.
One of our main results is that the non-diagonal 
(hereafter {\it skewed}) kinematics lead to a significant enhancement 
of the cross section. 
We present an estimate of the size of this effect, to
the leading-log accuracy with which the skewed splitting functions are known. 
At HERA energies $\ups(ns)$-states are produced at large effective scales 
(around 40, 60 and 75 GeV$^2$ for $n=1, 2, 3$, respectively) 
and at relatively high-$x$ (between $0.001$ and $0.02$). It follows  that 
the real part of the amplitude is large and we calculate it using a 
fit to the energy dependence of the  imaginary part 
and dispersion relations. 
Taking both effects into account leads to cross-sections which 
concur with the measured ones and rise very steeply with $W^2$,
the $\gamma P$ centre-of-mass energy (approximately  $W^{2(0.85)}$).

\FIGURE{
\setlength{\unitlength}{0.1bp}
\begin{picture}(2700,2700)(0,0)
\includegraphics{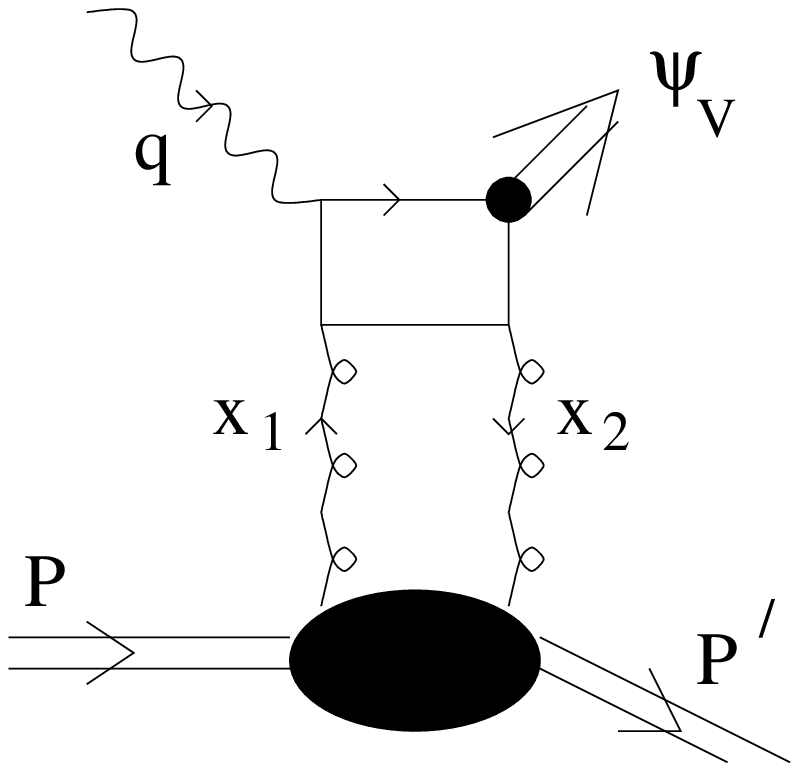}
\put(100,1230){%
}
\end{picture}
        \caption{Photoproduction of Heavy Vector mesons}    
    \label{figups}
}

This note is organised as follows. Firstly, we very briefly recap the relevant 
equations in \cite{FKS2}. 
Secondly, we discuss the rescaling procedure in detail. 
We then explain how the skewedness is calculated and implemented. 
Next, we give an explanation of the calculation of the real part of the amplitude. 
Finally, we present and discuss the resultant cross-sections, and conclude.

\section{Formulae for cross sections}

Hard exclusive processes factorize in QCD \cite{CFS} and at 
small-$x$ are driven by the 
exchange of two gluons in the $t$-channel. One such process is the diffractive 
electroproduction of heavy vector mesons, in which the exchanged gluons 
are connected to the $q {\bar q}$-pair fluctuation of the photon 
(virtuality $q^2 = - Q^2$) in the four possible ways 
(one of which is shown in fig.(\ref{figups})) 
and convoluted  with $\psi_V (z,k_t)$. 
The forward differential cross-section for a vector meson, mass $M_V$, 
containing a (current) quark of mass $m$, can be written as the product of an 
asymptotic expression and a finite $Q^2$ correction, 
${\cal C}(Q^2)$ (see \cite{FKS2} for more details) :

\begin{eqnarray}
\left. \frac{ d \sigma_{\gamma^{(*)} P \rightarrow V P} }{dt} \right|_{t=0} \, &=&  
 \frac{ 12\pi^3 \Gamma_V M_V^3 }{ \alpha_{em} (Q^2+4m^2)^4 } 
\times \nonumber \\
& & \, 
\left| \alpha_s (Q_{eff}^2) \left(1 + i \beta \right) xG_N(x,Q_{eff}^2) \right|^2 \,
\left(  1+\epsilon \frac{Q^2}{M_V^2} \right)  \,{\cal C}(Q^2) \, , \label{eqx1}
\end{eqnarray}
with 
\begin{equation}
{\cal C}(Q^2) = \left(\frac{\eta_V}{3} \right)^2 \,
\left( \frac{Q^2+4m^2}{Q^2+4m_{run}^2} \right)^4 \, T(Q^2) \, 
\frac{R(Q^2)+ \epsilon \frac{Q^2}{M_V^2} }{1 + \epsilon \frac{Q^2}{M_V^2} } \, ,  
\label{eqx2}
\end{equation}
\noindent where $\qf2 (Q^2), \Gamma_V, \epsilon$ are the 
effective transverse scale, 
the leptonic decay width and photon's polarisation respectively; 
$\beta$ is the ratio of real to imaginary parts of the amplitude.
The $k_t$-suppression factor $T$, $R$ and $\eta_V$ contain 
integrals involving the light-cone 
wavefunctions of the photon and vector meson. Hybrid wavefunctions are used 
which are designed to have the characteristic $z(1-z)$ behaviour at very small 
transverse distances expected from QCD (this implies $\eta_V = 3$).
This justifies the use of the QCD running mass 
and hence the correction for this in eq.(\ref{eqx2}).
Overall the use of the hybrid wavefunctions, in place of the pure 
non-relativistic wavefunctions, does not 
lead to significant effects in the cross sections.
In this paper we use the particular model of the quarkonium wavefunction due to 
Buchmueller and Tye \cite{BT}. They used a QCD-inspired potential, with a 
Coulomb piece corresponding to t-channel 
gluon exchange ($ \propto 1/r$) and a confinement piece ($\propto r$). 
The analysis produced a value of $m_b = 4.88 \,$ GeV for the quark mass. 
The significant QCD radiative corrections 
to the amplitude of heavy quarkonium transition into $q {\bar q}$-pair 
are accounted for by normalising the short-distance part of the hybrid 
light-cone wavefunctions to the  width of heavy quarkonium  decay into leptons.
The analysis of \cite{FKS2} indicated that once the short-distance corrections 
were included (for transverse distances less than $b_0 = 0.1 \, $ GeV$^{-1}$) 
several potential models produced similar results. As a result, 
we expect that the resultant model uncertainty is less than that 
from other sources and only consider the potential of \cite{BT}.

In the photoproduction limit we have
\begin{equation}
\sigma (\gamma P \rightarrow VP) =
 \frac{3 \pi^3 \, \Gamma_V M_V^3  (1 + \beta^2) }{ 64 \, 
\alpha_{EM} \, (m^2)^4 \, B_{D,V}}  
\, \left[ \, \alpha_s (Q_{eff}^2) \, xg (x,Q_{eff}^2) \, \right]^2 
\, {\cal C}(Q^2 = 0) \, , \label{eqx3}
\end{equation}

\noindent where, as usual, an exponential decay in $|t|$ is assumed,
with $B_{D,V}$ the slope parameter. 
Since these parameters have not yet been measured, we are forced to 
estimate them. The established trend is that the slope parameter 
decreases with increasing vector meson mass. This is a geometrical effect:  
most of $B_D$ comes from the proton end of the ladder where 
the typical transverse momenta are much smaller, 
the upper end contributes progressively less as the typical transverse 
momentum scales involved increase. On this basis
we expect $B_{D,\ups}$ to be a little less than that 
measured in photoproduction of $J/\psi$: 
$B_{D,J/\psi}= 4.4 \pm 0.3$ GeV$^{-2}$ (H1).
For the purpose of this paper we assume, for each state, 
the same value as that measured in electroproduction of $J/\psi$ by H1, 
$B_{D,V}= 3.9 $ GeV$^{-2}$ (see \cite{h1jpsi} and references therein), 
since the typical transverse scales at the upper vertices 
in these processes are similar. Of course, 
this introduces a further overall uncertainty in the 
normalisation of the cross sections of about $10-20 \%$.



\section{Rescaling: calculation of $\qf2$}


The procedure that we use  for determining the scale, $Q_{eff}^2$, 
in eq.(\ref{eqx3}) is similar to  that described in sec.(IIIa) of 
\cite{FKS1}. The amplitude the for photoproduction of  $\ups$ in 
co-ordinate space is given by:
\begin{equation}
A(\gamma_T P \rightarrow V_T P) \propto \int_0^1 \frac{dz}{z(1-z)} \int_0^{\infty} d^2 b_t 
\, m^2_{run} \, \psi_{\gamma,T} (z,b_t) \, {\hat \sigma}(b_t^2) \, \psi_{V,T} (z,b_t), 
\label{eqavm}
\end{equation}
\noindent where, 
\begin{equation}
{\hat \sigma}(b_t^2) = \frac{\pi^2}{3} \, b_t^2 \, \alpha_s (b_t^2) \, xg(x,b_t^2)
\label{eqshat}
\end{equation}
\noindent is the cross section for scattering a $q {\bar q}$ dipole
 of transverse size $b_t^2$ off the proton ($b_t$ is the conjugate variable to $k_t$). 
This quantity is expected to be universal. $\psi_{\gamma,T} (z,b_t) = K_{0} (b \, m_{run})$ and  $ \psi_{V,T} (z,b_t)$ are the light-cone wavefunctions of the transversely-polarised real photon and heavy vector meson in configuration space, respectively. 


In order to proceed we need to establish the relation between transverse sizes and
momentum scales, such as $Q^2_{eff}$, to determine the scale at which we must sample
the gluon density and $\alpha_s$ in eq.(\ref{eqshat}). We establish this relation
using the expression for the longitudinal structure function written in $b_t$-space: 
\begin{equation}
\sigma_L (x,Q^2) \propto \int dz \int_0^{\infty} d^2 b_t \, 
{\hat \sigma}(b_t^2) \, |\psi_{\gamma,L} (z,b_t)|^2 \, .
\label{eqsigl}
\end{equation}

\noindent We now assume that for a particular fixed $x,Q^2$ the product of 
transverse size and the momentum scale squared is constant, $b_t^2 \, Q^2 = \lambda$,
with $b_t^2$ measured in GeV$^{-2} = b_t^2$ (fm $^2$)/(hc)$^2$, with hc = 0.197 GeV fm.
We establish this constant from the typical or average transverse sizes 
contributing to the $b_t$-integral above. The average $b_t^2$ is defined 
to be that value up to which we must integrate in order to reach half of the full 
$b_t^2$ integral in eq.(\ref{eqsigl}). 
This average value is fed back into the integral via
$\lambda = <\!b_t^2\!> \, Q^2$. The integral is recalculated, sampling the gluon and 
$\alpha_s$  at $\lambda / <\!b_t^2\!>$ and a new median $b_t^2$ is found. 
The procedure is then iterated to convergence. 
It turns out that the resultant $\lambda$ depends weakly on $x, Q^2$. 
Table.(\ref{tablam}) shows 
$\lambda$ for a wide range of these kinematic variables 
(in \cite{FKS2} a constant value of $\lambda = 8.5$ at $x=10^{-3}$ 
is used for $J/\psi$).

\begin{table}[htbp]
\begin{center}
\begin{tabular}{|l|c|c|c|c|c|} \hline
 \, \, \, \, \, \, $Q^2$ & 10 & 30 & 50 & 70 & 90 \\ 
$x   $ &  &  &  &  &  \\ \hline
$1.0 \times 10^{-3}$ & 8.51  & 10.8 & 11.5  & 11.8  &12.0 \\
                     & 8.80  & 10.7 & 11.1  & 11.4  &11.7 \\ \hline
$5.0 \times 10^{-3}$ & 9.82  & 12.3 & 12.9  & 13.2  &13.4 \\
                     & 10.2  & 12.1 & 12.5  & 12.7  &12.9 \\ \hline
$9.0 \times 10^{-3}$ & 10.4  & 12.8 & 13.5  & 13.8  &13.9 \\
                     & 10.8  & 12.7 & 13.0  & 13.5  &13.6 \\ \hline
$1.3 \times 10^{-2}$ & 10.8  & 13.3 & 13.9  & 14.1  &14.3 \\
                     & 11.1  & 13.0 & 13.4  & 13.7  &13.8 \\ \hline
$1.7 \times 10^{-2}$ & 11.2  & 13.7 & 14.3  & 14.6  &14.7 \\
                     & 11.3  & 13.3 & 13.7  & 13.8  &14.0 \\ \hline
\end{tabular}
\end{center}
\caption{$\lambda (x,Q^2)$ as a function of $x$ and $Q^2$ (GeV$^2$). The upper values refer to CTEQ4L partons and the lower to MRSTLO.}
\label{tablam}
\end{table}

\begin{table}[htbp]
\begin{center}
\begin{tabular}{|l|c|c|c|c|c|c|} \hline
 &x & 0.001 &0.005 &0.009 &0.013 &0.017 \\ \hline
$\ups$ &CTEQ        & 37 & 40 & 40 & 40 & 41 \\
       &MRST        & 41 & 42 & 42 & 42 & 43 \\ \hline
$\Upsilon^{'}$ 	&CTEQ        & 58 & 61 & 62 & 63 & 65 \\
		&MRST        & 57 & 60 & 62 & 63 & 65 \\ \hline
$\Upsilon^{''}$ &CTEQ        & 71 & 76 & 77 & 80 & 80  \\
		&MRST        & 70 & 75 & 76 & 78 & 79  \\ \hline
\end{tabular}
\end{center}
\caption{The effective scale, $Q^2_{eff}$ in GeV$^2$,  
for each s-wave $\ups$-state as a function of $x$}
\label{tabqeff}
\end{table}

Now we have established the relation between transverse size and momentum scales 
we can use it to set the scale of  ${\hat \sigma} (b_t^2)$ 
in the $b_t$-integral in eq.(\ref{eqavm}). 
This allows us to establish the effective scale for the production of each state, by a
procedure similar to that used above. With some starting assumption for   
$Q^2_{eff}$ we use $\lambda (x,Q^2_{eff}) / b_t^2 $ as the scale 
at which $xg, \alpha_s, m^2_{run}$ are sampled in the integral. This requires 
some regulation at very small and very large values of $b_t^2$, however,  
the contributions to the overall integral from these regions are negligible.
We establish the median $b^2 =<\!b^2\!>$ for this integral, as above. 
This is then fed back in via $Q^2_{eff} (new) = \lambda (x,Q^2_{eff}(old))/<\!b^2\!>$ 
and the procedure is iterated to convergence.

In this way the value of the effective scale depends on the dominant values of $b^2_t$.
Dipoles in the photon which most closely correspond to the relevant s-wave state,  
(have the ``right transverse size'' and momentum sharing $z$) contribute most to 
the production. The rescaling procedure is designed to reflect this.
The scale varies depending on the state in question 
since the relevant light-cone wavefunctions weight 
the integral in eq.(\ref{eqavm}) differently.
The resulting effective scale, $Q_{eff}^2$, depends weakly on $x$ and strongly on the state concerned (see table.(\ref{tabqeff})).

The fact that $Q^2_{eff}$ increases with the mass of the state agrees
with \cite{FKS1,FKS2} and is perhaps counter-intuitive given that the quarks 
are less tightly bound in the higher s-wave states, leading to 
larger typical sizes. 
It is worth remembering, however, that the precise 
momentum scale is governed by a convolution  of the photon 
and vector meson light-cone wavefunctions, and that higher 
states also contain nodes which further confuses the issue 
from an intuitive point of view. In fact, the typical transverse momenta 
contributing to the production of these higher mass states is larger
(see table.(\ref{tabktz})) and this is reflected in the larger $Q^2_{eff}$. 
The precise values of this effective scale are sensitive to the details 
of the vector meson wavefunction. The outlined rescaling procedure represents 
a reasonable estimate given the current knowledge of this quantity.

\section{Calculation of the effect of skewedness}

Recently, there has been considerable progress 
in calculating and understanding skewed parton distributions, 
which probe new non-perturbative information about hadrons and are a 
generalisation of conventional parton distributions  
(for an extensive list of references see \cite{JI}). 
The former replace the latter in  
expressions for hard exclusive cross-sections. 
They are relevant for a wide range of hard exclusive
processes such as deeply-virtual Compton scattering, 
diffractive dijets in photoproduction, and most importantly for our purposes, 
photo- and electroproduction of vector mesons.

In the latter processes, the skewedness, $\delta$, arises 
from the need to convert a 
space-like $q^2 = - Q^2$, 
into a time-like $M_V^2$, 
and is given by the difference in momentum fractions 
carried by the outgoing ($x_1$) 
and returning ($x_2$) gluons (see fig.(\ref{figups})),
\begin{equation}
x_1 =  \frac {M_{q {\bar q}}^2 + Q^2 }{W^2 + Q^2} \, , \,    
x_2 = \frac{M_{q {\bar q}}^2 - M_V^2}{W^2 + Q^2}  ,
\end{equation}
\begin{equation}
\delta = x_1 - x_2  =  \frac{M_V^2 + Q^2}{W^2 + Q^2} , 
\end{equation}

\noindent where $M^2_{q {\bar q}}$ is the mass of the intermediate $q {\bar q}$-state.
In terms of the light-cone variables it is given by
\begin{equation}
M^2_{q {\bar q}}  =  \frac{k_t^2 + m_q^2}{z(1-z)}.
\label{eqmqq}
\end{equation}

\noindent For photoproduction of $\ups$ at HERA, $\delta = M_{\ups}^2 / W^2$ 
is small and lies in the range $\{0.0011,0.017\}$, 
although it is an order of magnitude larger than
 for $J/\psi$-photoproduction in the same energy range.

It was argued in \cite{FG1} that for small $x$ and 
for $Q^2_0$, where parton densities weakly depend on
$x$, the dependence on $\delta$ can be neglected. 
It was further demonstrated in \cite{FG1} that for large $Q^2$ 
and small $x$ the main contribution to the skewed parton densities 
comes from the parton densities at $Q^2_0$ scale where  
$\tilde{x}\gg x$, for  which the skewedness is very small. Hence 
the answer for these kinematics does not depend on this approximation.
Freund and Guzey \cite{FG2} have modified the DGLAP-evolution \cite{DGLAP} 
package of the CTEQ collaboration \cite{CTEQ}, which is based on a
numerical grid integration, to produce the skewed gluon functions,
$G_{\delta}(x_1,Q^2)$, for any $x_1$ at a fixed value of  $\delta$ 
(it is straightforward to write an interpolating subroutine to obtain 
$G$ for any $\delta$).  We use this code and the 
approximation that the skewed and
conventional distributions are the same at the starting scale, $Q_0^2$.

  
It is then sufficient to replace the conventional splitting 
functions with their skewed generalisations in the evolution 
(i.e.  $P_{ab} (x) \rightarrow P_{ab} (\delta,x)$). 
Unfortunately the latter are only known to leading-log accuracy at present 
(although, very recently progress has been reported \cite{belit} on the 
next-to-leading order). 
For consistency one is forced to use only leading-log parton distributions 
at the input scale; we use the two most recent leading-order distributions, i.e. 
CTEQ4L \cite{cteqlo} and MRSTLO\footnote{We thank R. Roberts for providing the parameters at the starting scale.} 
\cite{mrstlo}, and evolve to leading-log($Q^2$) accuracy.
The code produces the skewed gluon distribution $G_{\delta}(x_1,Q^2_{eff})$ 
to replace the ordinary gluon density, 
$xg(x,Q^2_{eff})$ of eq.(\ref{eqx3}). 
The two distributions are shown in fig.(\ref{figglue}), in the relevant $x$-range
for $Q^2_{eff} = 40 $ GeV$^2$ characteristic of $\ups(1s)$-production 
(see table.(\ref{tabqeff})). 
This replacement leads to an overall enhancement factor of about $(1.6)^2 \simeq 2.6$ 
for $\ups$(1s,2s,3s) cross-sections.

\FIGURE{
\caption{A comparison of skewed and conventional gluon densities at a scale of 
$Q^2 = 40$ GeV$^2$, the effective scale for $\ups$-photoproduction. Conventional 
leading-order input distributions are used in both cases (at starting scales 
$Q_0^2 = 1.6$ and $1.0 $ GeV$^2$ for CTEQ4L \cite{cteqlo} and MRSTLO \cite{mrstlo},
  respectively). The skewed distributions are the upper curves  and have fixed 
$x_1/\delta = 1.2$.}
\setlength{\unitlength}{0.1bp}
\begin{picture}(3600,2160)(0,0)
\includegraphics{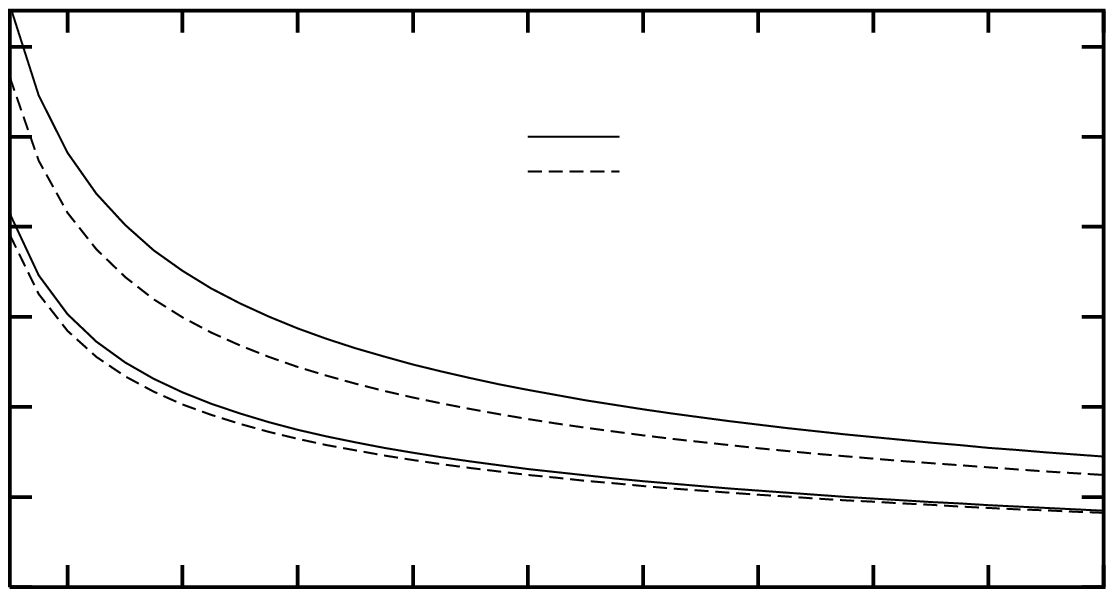}
\put(1842,1597){\makebox(0,0)[r]{cteq}}
\put(1842,1697){\makebox(0,0)[r]{mrst}}
\put(1975,150){\makebox(0,0){$\Large x, x_1$}}
\put(100,1230){%
\makebox(0,0)[b]{\shortstack{$\Large xg(x,40), G_{\delta} (x_1,40)$}}%
}
\put(3550,300){\makebox(0,0){0.02}}
\put(3218,300){\makebox(0,0){0.018}}
\put(2887,300){\makebox(0,0){0.016}}
\put(2555,300){\makebox(0,0){0.014}}
\put(2224,300){\makebox(0,0){0.012}}
\put(1892,300){\makebox(0,0){0.01}}
\put(1561,300){\makebox(0,0){0.008}}
\put(1229,300){\makebox(0,0){0.006}}
\put(897,300){\makebox(0,0){0.004}}
\put(566,300){\makebox(0,0){0.002}}
\put(350,1956){\makebox(0,0)[r]{30}}
\put(350,1697){\makebox(0,0)[r]{25}}
\put(350,1438){\makebox(0,0)[r]{20}}
\put(350,1178){\makebox(0,0)[r]{15}}
\put(350,919){\makebox(0,0)[r]{10}}
\put(350,659){\makebox(0,0)[r]{5}}
\put(350,400){\makebox(0,0)[r]{0}}
\end{picture}
\label{figglue}
}

It is necessary to determine the appropriate value of $x_1$ to use.
We estimate this by calculating the average values of transverse
 momentum and light-cone momentum sharing $<\!k_t\!>$ and 
$<\!z\!>$ contributing to $M_{q {\bar q}}^2$ in the relevant loop integral.
A more accurate calculation would integrate explicitly over $M_{q {\bar q}}^2$, but in 
this case all direct connection to the measured forward distribution, 
and hence predictability, would be lost (until the input densities for 
skewed evolution have been explicitly measured themselves).

The amplitude of eq.(\ref{eqavm}) is used to calculate the relevant value of $M_{q {\bar q}}^2$  for the particular value of $W$. 
First of all we calculate $<\!\!z\!\!>$. 
We then integrate over z, take a Fourier transform back to 
$k_t$-space and calculate the median value of the $k_t$-integral by the 
method outlined above for calculating medians of integrals. Both $<\!z\!>$ and 
$<\!k_t^2\!>$ are then fed into eq.(\ref{eqmqq}) to determine $M_{q {\bar q}}^2$. 
The range of values of $<\!k_t^2\!>$ and typical $<\!z\!>$ 
for each s-wave state are shown in table.(\ref{tabktz}).
The result depends weakly on energy, $W$. 
Table.(\ref{tabx1od}) shows the resultant values of 
$x_1/\delta = M_{q {\bar q}}^2/M_V^2$ as a function of $W$ for the 
three s-wave states. The code for calculating the cross sections 
automatically takes this variation into account.

Note that as soon as $x_1/\delta$ is small we are not sensitive to 
a specific value of this ratio since in this limit 
the ratio of skewed and diagonal
densities  weakly depends on $x_1/\delta$. 
This justifies the use of the conventional distributions in 
${\hat \sigma}$ when calculating $ M_{q {\bar q}}^2$.

\begin{table}[htbp]
\begin{center}
\begin{tabular}{|l|c|c|c|c|} \hline
       		&      	& $<k_t^2>$ (80 GeV) &  $<k_t^2>$ (280 GeV) & $<z>$  \\ \hline
$\ups$ 		&CTEQ  	& 4.04               & 5.34                 & 0.40   \\
       		&MRST  	& 4.54               & 5.90                 & 0.40   \\ \hline
$\Upsilon^{'}$ 	&CTEQ 	& 4.88               & 6.20                 & 0.28   \\
		&MRST 	& 5.43 		     & 6.71                 & 0.28   \\ \hline
$\Upsilon^{''}$ &CTEQ 	& 5.90               & 7.34                 & 0.26   \\
		&MRST 	& 6.50               & 7.78                 & 0.26   \\ \hline
\end{tabular}
\end{center}
\caption{Averages of $k_t^2$ and $z$ for each s-wave $\ups$-state.}
\label{tabktz}
\end{table}

\begin{table}[htbp]
\begin{center}
\begin{tabular}{|l|c|c|c|c|c|c|} \hline
&$W$ (GeV)       	& 80 & 130 & 180 & 230 & 280 \\ \hline
$\ups$ 		&CTEQ  & 1.14 & 1.16 & 1.17 & 1.18 & 1.19 \\
       		&MRST  & 1.17 & 1.20 & 1.21 & 1.22 & 1.24 \\ \hline
$\Upsilon^{'}$ 	&CTEQ & 1.23 & 1.26 & 1.28 & 1.29 & 1.30 \\
		&MRST & 1.28 & 1.30 & 1.32 & 1.33 & 1.34 \\ \hline
$\Upsilon^{''}$ &CTEQ & 1.28 & 1.30 & 1.32 & 1.33 & 1.34 \\
		&MRST & 1.32 & 1.35 & 1.36 & 1.37 & 1.38 \\ \hline
\end{tabular}
\end{center}
\caption{The ratio $x_1/\delta$ for each s-wave $\ups$-state as a function of $W$.}
\label{tabx1od}
\end{table}

\section{Calculating the real part of the amplitude}

It is convenient  to use  the dispersion representation in energy to calculate
the real part of the scattering amplitude. In doing so we automatically take
into account a possible contribution of the region of $x_1>0, x_2<0$ in the
real part. 
Previously, the ratio of real to imaginary parts of the amplitude,  $\beta$,
was calculated using the approximate solution of this dispersion relation:
\begin{equation}    
\beta = \frac{\pi}{2} \frac{d \ln (xg(x,Q^2))}{d \ln (1/x) }  \, ,
\label{eqx5}
\end{equation}
\noindent first derived in \cite{GM}, and used in \cite{FKS1,FKS2,RRML}. This is appropriate 
at small $x$ and fairly low scales where the gluon density and hence the 
imaginary part of the amplitude may be approximated by a single power in energy 
($Im A \propto xg/x \propto (W^2)^{\alpha}$). 
The result is then $\beta = \frac{\pi}{2} (\alpha -1)$, for
$\alpha$ sufficiently close to 1.


\begin{table}[htbp]
\begin{center}
\begin{tabular}{|c|c|c|c|c|c|c|c|} \hline
 & \multicolumn{2}{|c|}{$\ups$} &  \multicolumn{2}{|c|}{$\ups^{'}$} &
\multicolumn{2}{|c|}{$\ups^{''}$} & $\ups$ (eq.(\ref{eqx5})) \\ \hline
$W$ (GeV) & CTEQ & MRST & CTEQ & MRST & CTEQ & MRST & CTEQ\\ \hline
80        & 1.25 & 1.27 & 1.55 & 1.56 & 1.78 & 1.78 & 0.88\\ \hline
130       & 0.76 & 0.79 & 0.89 & 0.91 & 0.94 & 0.99 & 0.68\\ \hline
180       & 0.62 & 0.65 & 0.71 & 0.73 & 0.72 & 0.78 & 0.60\\ \hline
230       & 0.55 & 0.58 & 0.63 & 0.64 & 0.63 & 0.68 & 0.56\\ \hline
280       & 0.51 & 0.54 & 0.58 & 0.59 & 0.58 & 0.63 & 0.53\\ \hline
\end{tabular}
\end{center}
\caption{Relative contribution of the real part of the amplitude, $\beta^2$, as a function of $W$ for photoproduction of $\ups$(1s), $\ups$(2s), $\ups$(3s). For comparison the last column shows the values of $\beta^2$ obtained for $\ups$ using eq.(\ref{eqx5}), sampling the CTEQ4L gluon at $Q^2_{eff} = 40$GeV$^2$.}
\label{tabbeta}
\end{table}

At the larger values of $x$ and higher scales, 
$Q^2_{eff}$, relevant to photoproduction of $\ups$-states it is necessary to include an additional sub-leading power in $1/x$ or equivalently in $W^2$.
For fixed $x_1/\delta$ (we use the values at $W=180$ GeV, see 
table.(\ref{tabx1od})) and $Q^2_{eff}$ the skewed gluon distribution 
is a function of $W^2$ only. We perform a two-power Regge-type fit to this 
distribution,  at the relevant fixed effective scale, 
over the whole range in $W^2$ using MINUIT \cite{minuit}:
\begin{equation}
G_{\delta} (x_1, Q^2_{eff}) = a W^{2b} + c  W^{2d} \, . \label{eqx6}
\end{equation}

We may then use analyticity of the amplitude directly to 
construct a dispersion relation and hence determine $\beta$, 

\begin{equation}
\beta (W) = \frac{Re A}{Im A} =  - \frac {a W^{2(b+1)} \cot(\frac{\pi (b+1)}{2}) 
+ c  W^{2(d+1)} \cot(\frac{\pi (d+1)}{2}) }{a W^{2(b+1)} + c W^{2(d+1)}}
 \, . \label{eqx7}
\end{equation}

\noindent Table.(\ref{tabbeta}) shows how $\beta^2$ changes as a function of $W$, 
these values are used in the cross sections of eq.(\ref{eqx3}). For a comparison, 
we also show the values of $\beta^2$ for $\ups$ using eq.(\ref{eqx5}). 
As expected these differ most for smaller values of $W$.

\section{Results}

The cross section for photoproduction of the $\ups$ states is
\begin{equation}
\sigma (\gamma P \rightarrow VP) =
 \frac{3 \pi^3 \, \Gamma_V M_V^3  (1 + \beta^2) }{ 64 \, 
\alpha_{em} \, (m^2)^4 \, B_{D,V}}  
\, \left[ \, \alpha_s (Q_{eff}^2) \, g_{\delta} (x_1,Q_{eff}^2) \, \right]^2 
\, {\cal C}(Q^2 = 0) \, . 
\label{eqsigf}
\end{equation}
\noindent The effective scale, $\qf2$, $x_1/\delta$ and $\beta^2$ all depend on 
the state concerned and weakly on energy as indicated in the tables.(2,4,5) above.
The overall suppression factor, 
$C(Q^2=0)$ is given by
\begin{equation}
C (Q^2=0) = (\frac{m^2}{m_{run}^2})^4 \left[
\frac{m_{run}^6}{M_V^2} \frac
{\int \frac{dz}{z(1-z)} \int db \, b^3 \, \psi_{V,T} (z,b) \, \psi_{\gamma} (z,b)}
{\int \frac{dz}{z(1-z)} \, \psi_{V,T} (z,b=0) } \right]^2 \, .
\end{equation}

The resolution of the muon chambers at the HERA experiments is 
such that the three contributing s-wave states 
$\ups ,\ups^{'} ,\ups^{''}$ are not resolved \cite{DATA}. 
Hence, what is actually measured is the sum of the  
production rates of the mesons times their respective branching 
fractions to muons (Br($\ups$ (ns) $\rightarrow \mu^+ \mu^- $) = 
2.48, 1.31, 1.81 \% for $n=1,2,3$, respectively \cite{PDG}). 
In fig.(\ref{figsum}) we present our predictions for this measured quantity 
using MRST (upper solid curve) and CTEQ (lower solid curve) input parton distributions. 
This illustrates the agreement of the model with the data collected so far. 
Also shown (dashed curves) are the predictions for $\ups (1s)$ alone for the 
two input distributions. There are two reasons why the MRST curves should be higher than 
CTEQ curves. Firstly, the skewed parton distributions are higher 
(see Fig.(\ref{figglue})). Secondly, smaller values for $\Lambda_{QCD}$ 
are used in MRSTLO than in CTEQ4L, this makes the value of $\alpha_s$ 
larger at a given effective scale. 

\FIGURE{
\setlength{\unitlength}{0.1bp}
\begin{picture}(3600,2160)(0,0)
\includegraphics{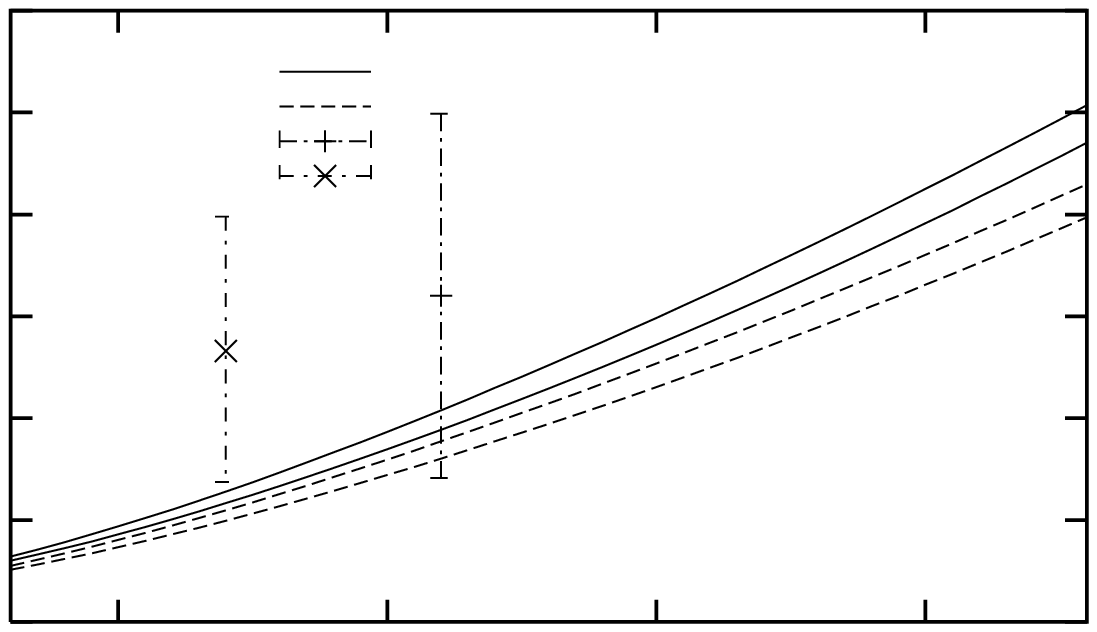}
\put(1075,1584){\makebox(0,0)[r]{ZEUS}}%
\put(1075,1684){\makebox(0,0)[r]{H1}}%
\put(1075,1784){\makebox(0,0)[r]{1s only}}%
\put(1075,1884){\makebox(0,0)[r]{All s-states}}%
\put(1900,50){\makebox(0,0){W (GeV)}}%
\put(100,1180){%
\makebox(0,0)[b]{\shortstack{$\Sigma_i  \, \sigma_i$ .  B$_i$$(\Upsilon \rightarrow \mu^+ \mu^-)$ (pb)}}%
}%
\put(2985,200){\makebox(0,0){250}}%
\put(2210,200){\makebox(0,0){200}}%
\put(1435,200){\makebox(0,0){150}}%
\put(660,200){\makebox(0,0){100}}%
\put(300,2060){\makebox(0,0)[r]{30}}%
\put(300,1767){\makebox(0,0)[r]{25}}%
\put(300,1473){\makebox(0,0)[r]{20}}%
\put(300,1180){\makebox(0,0)[r]{15}}%
\put(300,887){\makebox(0,0)[r]{10}}%
\put(300,593){\makebox(0,0)[r]{5}}%
\put(300,300){\makebox(0,0)[r]{0}}%
\end{picture}
    \caption{Sum of cross sections times branching ratios to muons for $\ups$(1s), $\ups^{'}$(2s), $\ups^{''}$(3s) as a function of energy. The dashed curves are for $\ups $(1s) alone. The upper (lower) curves in each case correspond to MRSTLO (CTEQ4L) partons at the starting scale. Also shown are the data from ZEUS and H1 (preliminary), at their respective mean energies, with systematic and statistical errors added in quadrature.}
    \label{figsum}}

Fig.(\ref{figsig}) shows our prediction for the photoproduction of $\ups$(1s) 
as function of energy in the HERA range, using both input distributions.
 A very steep rise is expected, as can been seen from the figure, 
this corresponds approximately to $W^{1.7}$ over the range shown, 
i.e. almost a full power in $W$ stronger than that seen in 
$J/\psi$ production (a recent fit\cite{h1jpsi} to ZEUS and H1 data revealed a power of
$ 0.8 \pm 0.1$). This very steep rise is due to the sampling of the gluon 
at the large scale,  $Q^2_{eff} \simeq 40 $ GeV$^2$, where it is rising steeply with 
energy. In practice the steepness of the observed rise may prove to be a useful way 
to discriminate between models which have rescaling and those that do not, 
since for a fixed range in $x$ the steepness increases with scale.

\FIGURE{
\setlength{\unitlength}{0.1bp}
\begin{picture}(3600,2160)(0,0)
\includegraphics{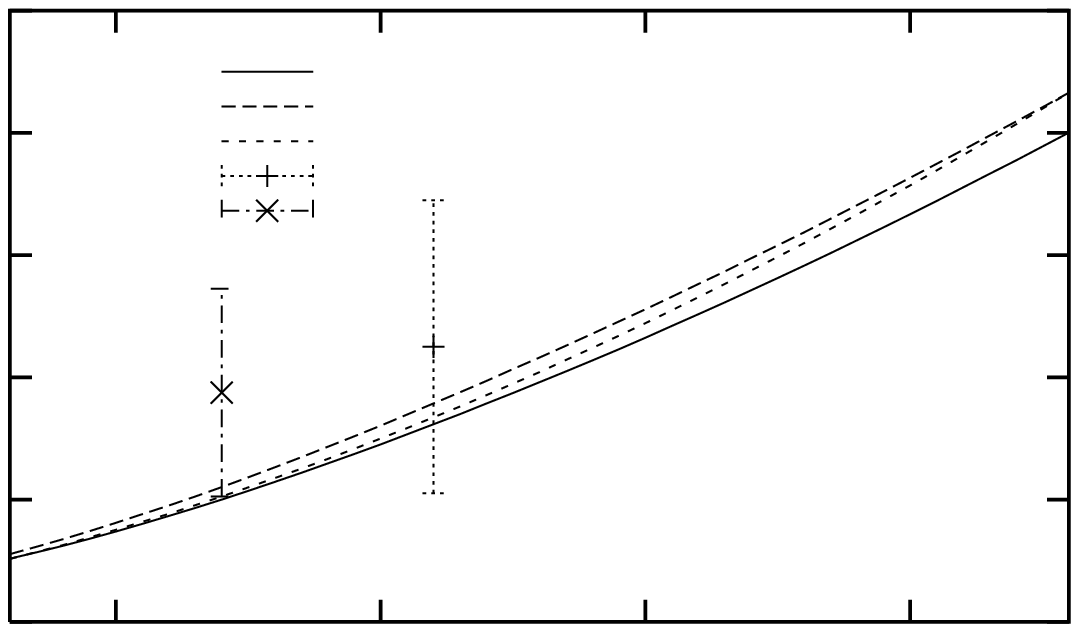}
\put(960,1484){\makebox(0,0)[r]{ZEUS}}%
\put(960,1584){\makebox(0,0)[r]{H1}}%
\put(960,1684){\makebox(0,0)[r]{$A W^{1.7}$}}%
\put(960,1784){\makebox(0,0)[r]{mrst}}%
\put(960,1884){\makebox(0,0)[r]{cteq4l}}%
\put(1925,50){\makebox(0,0){W (GeV)}}%
\put(100,1180){%
\makebox(0,0)[b]{\shortstack{$\sigma (\gamma $P$ \rightarrow \Upsilon $P)  (nb)}}%
}%
\put(2993,200){\makebox(0,0){250}}%
\put(2230,200){\makebox(0,0){200}}%
\put(1468,200){\makebox(0,0){150}}%
\put(705,200){\makebox(0,0){100}}%
\put(350,2060){\makebox(0,0)[r]{1}}%
\put(350,1708){\makebox(0,0)[r]{0.8}}%
\put(350,1356){\makebox(0,0)[r]{0.6}}%
\put(350,1004){\makebox(0,0)[r]{0.4}}%
\put(350,652){\makebox(0,0)[r]{0.2}}%
\put(350,300){\makebox(0,0)[r]{0}}%
\end{picture}
    \caption{Cross section for photoproduction of $\ups$(1s) in the HERA energy range, compared to values quoted by the HERA experiments \cite{DATA} at their respective mean energies. An additional curve, $AW^{1.7}$, with $A$ normalised to the CTEQ4L cross section at $W=80$ GeV,  is also shown to indicate the very steep rise with energy.}
\label{figsig}
}

\FIGURE{
\setlength{\unitlength}{0.1bp}
\begin{picture}(3600,2160)(0,0)
\includegraphics{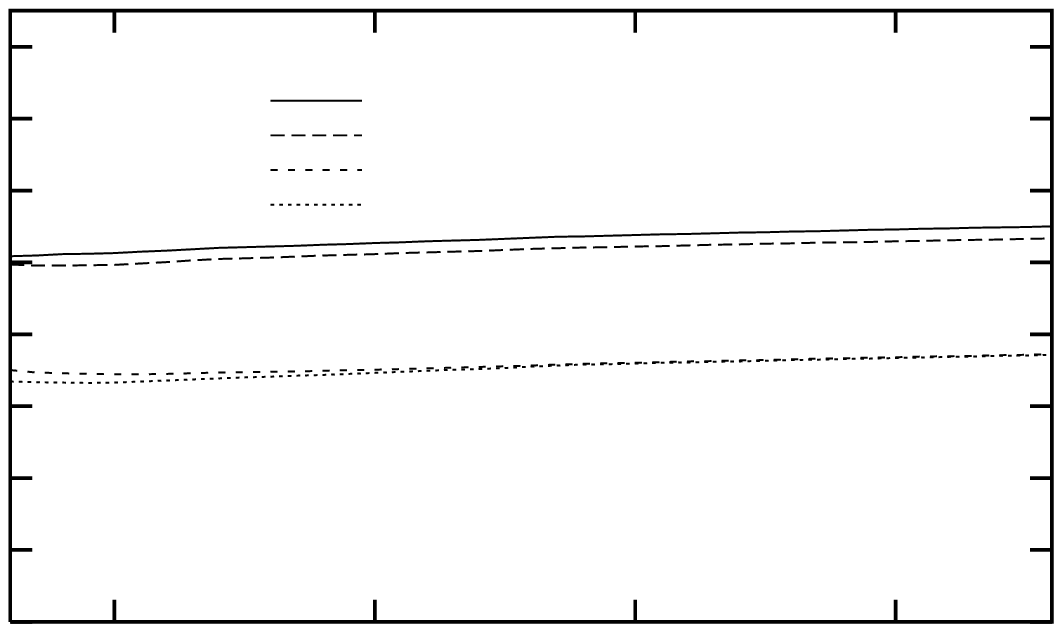}
\put(1150,1501){\makebox(0,0)[r]{$R_{3s}$ mrst}}%
\put(1150,1601){\makebox(0,0)[r]{$R_{3s}$ cteq}}%
\put(1150,1701){\makebox(0,0)[r]{$R_{2s}$ mrst}}%
\put(1150,1801){\makebox(0,0)[r]{$R_{2s}$ cteq}}%
\put(1950,50){\makebox(0,0){W (GeV)}}%
\put(100,1180){%
\makebox(0,0)[b]{\shortstack{$R_{2s},R_{3s}$}}%
}%
\put(3000,200){\makebox(0,0){250}}%
\put(2250,200){\makebox(0,0){200}}%
\put(1500,200){\makebox(0,0){150}}%
\put(750,200){\makebox(0,0){100}}%
\put(400,1956){\makebox(0,0)[r]{0.16}}%
\put(400,1749){\makebox(0,0)[r]{0.14}}%
\put(400,1542){\makebox(0,0)[r]{0.12}}%
\put(400,1335){\makebox(0,0)[r]{0.1}}%
\put(400,1128){\makebox(0,0)[r]{0.08}}%
\put(400,921){\makebox(0,0)[r]{0.06}}%
\put(400,714){\makebox(0,0)[r]{0.04}}%
\put(400,507){\makebox(0,0)[r]{0.02}}%
\put(400,300){\makebox(0,0)[r]{0}}%
\end{picture}
    \caption{Ratios of cross-sections times branching ratio to muons: $R_{2s} 
= \sigma B (2s) / \sigma B (1s)$, $R_{3s} = \sigma B (3s) / \sigma B (1s)$.}
    \label{figrat}
}

Figure.(\ref{figrat}) shows predictions for ratios of the products of 
cross-sections and branching ratios to muons, for the 
$\ups^{'} ,\ups^{''}$ relative to $\ups$ as a function of energy.
The different $k_t$-suppression factors for the different states 
(for CTEQ4L   $C(Q^2=0) = 0.290, 0.139, 0.0820$, for 
MRSTLO  $C(Q^2=0) = 0.283, 0.138, 0.0817$, respectively) 
imply that the relative rates of 
the three states are not expected to be the same as those 
found at  CDF \cite{CDF}, and used by ZEUS and H1 to unfold 
and produce a cross section for $\ups$(1s) production.  The CDF data implies 
that $\ups$(1s) is responsible for about $70\%$ of the signal 
whereas our calculation indicates a larger value of  
about $85\%$ (see also fig.(\ref{figsum})). For this reason a comparison of the  
predictions in Fig.(\ref{figsig}) with the values quoted by the experiments 
(also shown), must be done with care.

Although the states cannot be separated this predominance of $\ups$(1s) 
should reflect itself in the average observed value of the mean mass of 
the muon pair in the signal.

\section{Conclusions and discussion}

We have presented predictions for the cross sections of diffractive photoproduction of 
$\ups$-states in the HERA range, based on a previous analysis\cite{FKS2}. 
In this paper two new effects, the off-diagonal nature of the amplitude and the 
largeness of it's real part are found to be important and lead to  a significant 
enhancement of the normalisation of the cross sections. We perform a reanalysis of
rescaling procedure used to set the scale of the production of the 
three s-wave states. We find good agreement with the first data from HERA. 
One striking prediction of the analysis is a very strong rise of the 
cross sections with energy which is driven by the skewed gluon density (see 
eq.(\ref{eqsigf}) and fig.(\ref{figglue}))

Throughout the paper we have used the hybrid light-cone wavefunctions suggested in 
\cite{FKS2}, which were constructed to agree with quarkonium wavefunctions at 
large distances and QCD at small distances. These wavefunctions have  
significant Fermi motion of the quarks inside the bound states. Gauge invariance 
demands that if one has finite-$k_t$ one also needs gluonic degrees of freedom. 
So far only the contribution to the cross section of
the lowest order Fock states of the vector meson and photon
($|b {\bar b}\!>$) have been considered. 
The next step is to calculate higher order 
Fock states which also include the gluon degrees of freedom. The 
connection between these light-cone wavefunctions and quarkonium wavefunctions 
derived from solving the Schr\"{o}dinger equation for a particular potential 
remains an open question, which such a calculation would begin to address.

\acknowledgments 

The authors gratefully acknowledge Werner K\"{o}pf, Andreas Freund and 
Vadim Guzey for providing and explaining various computer codes.
We also thank the referee for his valuable suggestions.
M.M. is happy to acknowledge PPARC for financial assistance 
(grant number: GR/L56244). M.S. would like to thank DESY for hospitality 
during the time this work was done. 
The work of M.S and L.F. is supported by the U.S. Department of Energy and BSF.


\end{document}